\documentclass[aps,twocolumn,prl,groupedaddress,showpacs]{revtex4-1}

\usepackage{graphicx,float}
\usepackage[normalem]{ulem}
\usepackage{color}

\newcommand{\Niso}{$^{14}$N }
\newcommand{\ket}[1]{|#1\rangle}

\begin{document}

\title{High fidelity transfer and storage of photon states in a single nuclear spin}

\author{Sen Yang$^1$$^{\S}$}
\email{s.yang@physik.uni-stuttgart.de}
\author{Ya Wang$^1$$^{\S}$}
\author{D. D. Bhaktavatsala Rao$^1$}
\author{Thai Hien Tran$^1$}
\author{S. Ali Momenzadeh$^1$}
\author{Roland Nagy$^1$}
\author{M. Markham $^2$}
\author{D. J. Twitchen $^3$}
\author{Ping Wang $^4$}
\author{Wen Yang $^4$}
\author{Rainer St\"{o}hr$^{1,5}$}
\author{Philipp Neumann$^1$}
\author{Hideo Kosaka$^6$}
\author{J\"{o}rg Wrachtrup$^1$}
\email{wrachtrup@physik.uni-stuttgart.de}

\affiliation{$^1$ 3. Physikalisches Institut, Research Center SCOPE, and MPI for Solid State Research, University of Stuttgart, Pfaffenwaldring 57, 70569 Stuttgart, Germany}
\affiliation{$^2$ Element Six Ltd, Global Innovation Centre, Fermi Avenue, Harwell Oxford, Didcot, Oxfordshire, OX11 0QR, UK}
\affiliation{$^3$ Element Six Technologies US Corporation, 3901 Burton Drive, Santa Clara, CA 95054, USA}
\affiliation{$^4$ Beijing Computational Science Research Center, Beijing 100084, China}
\affiliation{$^5$Institute for Quantum Computing, University of Waterloo, N2L 3G1, Canada}
\affiliation{$^6$ Department of Physics, Faculty of Engineering, Yokohama National University, Tokiwadai, Hodogayaku, Yokohama, 240-8501, Japan}
\affiliation{$^{\S}$ These authors contributed equally to this work}

\begin{abstract}
\end{abstract}

\pacs{}
\maketitle

\textbf{Building a quantum repeater network for long distance quantum communication requires photons and quantum registers that comprise qubits for interaction with light, good memory capabilities and processing qubits for storage and manipulation of photons. Here we demonstrate a key step, the coherent transfer of a photon in a single solid-state nuclear spin qubit with an average fidelity of $98\%$ and storage over 10 seconds. The storage process is achieved by coherently transferring a photon to an entangled electron-nuclear spin state of a nitrogen vacancy centre in diamond, confirmed by heralding through high fidelity single-shot readout of the electronic spin states. Stored photon states are robust against repetitive optical writing operations, required for repeater nodes. The photon-electron spin interface and the nuclear spin memory demonstrated here constitutes a major step towards practical quantum networks, and surprisingly also paves the way towards a novel entangled photon source for photonic quantum computing.}

A quantum repeater network is intended to distribute entanglement between distant nodes realizing an elementary quantum network \cite{BDCZ}. Building up such a network requires photon sources (single or entangled pairs), processing nodes with the ability to make (i) optical or spin Bell-state measurements, (ii) long coherence times and (iii) ability for entanglement purification or quantum-error correction \cite{RevModPhys_repeater, Kimble_internet} . With such strong requirements it is hard to find physical systems meeting all of the above criteria. In this regard ensembles of atomic gases, trapped ions and solid state systems are intensively studied ~\cite{BDCZ,DLCZ,RevModPhys_interface,RevModPhys_repeater,Pan_repeater,Gisin_storage}. While e.g., atomic systems provide high interaction efficiency with photons, rare earth systems on the other hand show long coherence times, all are required for processing the absorbed/emitted photons. As opposed to ensembles, single particles though typically have a significantly less interaction efficiency with photons, however are useful for quantum networks due to their ability for in situ information processing~\cite{RevModPhys_duan,Rempe_atom_memory,Rempe_atom_network,Rempe_PRL},
like entanglement purification~\cite{Bennett_purification,Wineland_purification}.

\begin{figure*}
\includegraphics[width=0.95\textwidth]{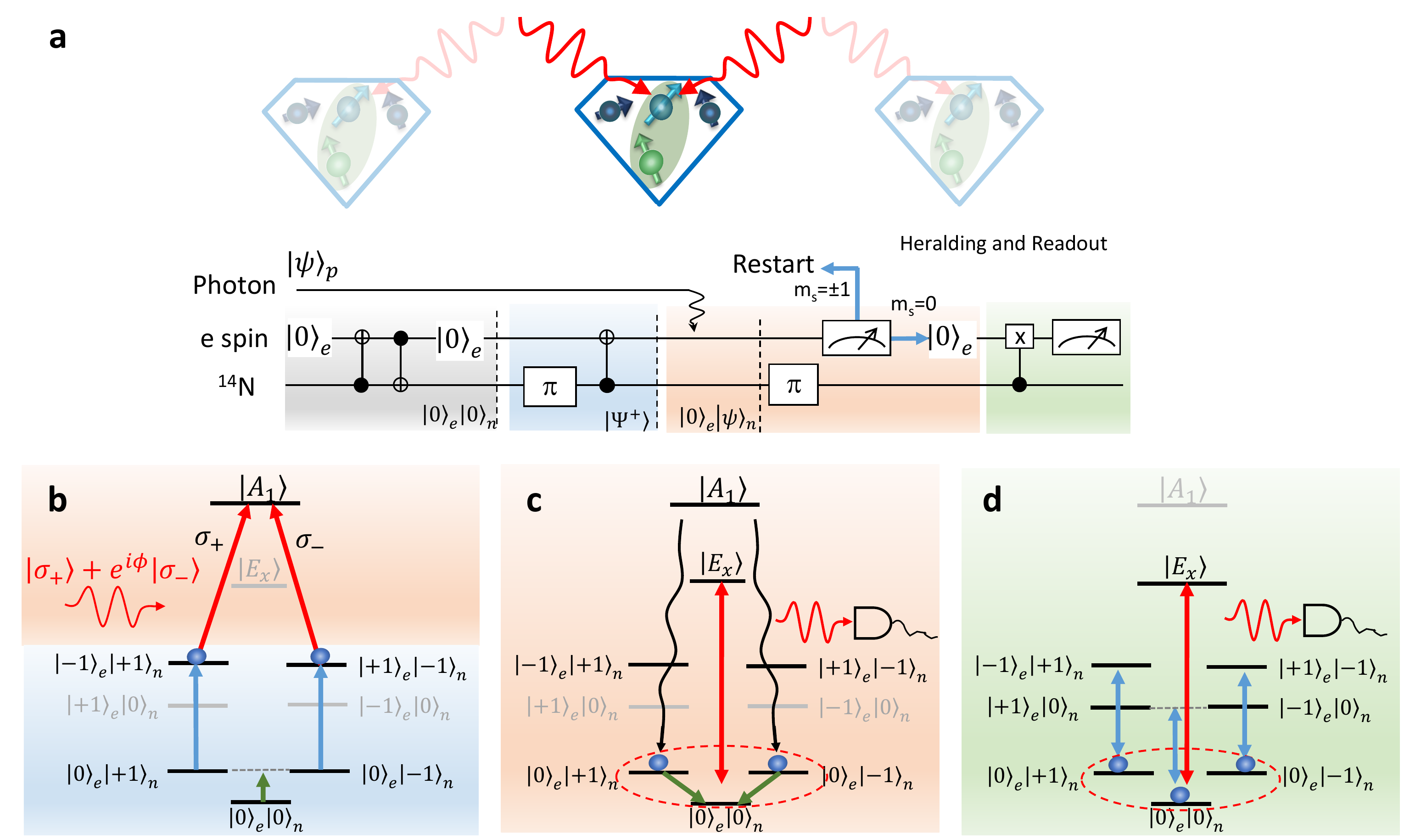}
	\caption{\textbf{Quantum interface connecting light to single nuclear spin in NV.}
 {\textbf{(a)(upper)}}, Schematic representation of an absorption based quantum repeater network with multiple nodes made of solid-state spins in diamond interacting with photons that are generated from entangled photon sources that transmit photons to adjacent quantum nodes thereby entangling them. Each node consists of one communicator qubit (electron) and few memory qubits (nuclear spins). The initial electron-nuclear spin entanglement followed by electron-photon entanglement leads to the storage of the photon in the nuclear spin resembling photon-nuclear spin teleportation. Multiple operations on the communicator spin (here symbolized by two impinging photons) without compromising coherence stored in the nuclear spin is an essential requirement for the scheme. {\textbf{(lower)}},  Circuit diagram and pulse sequence for storing a photon into the nuclear spin and its verification.
  {\textbf{(b) (lower)}}, The Bell-state preparation of the electron-nuclear spin system from the initial state $\ket{0}_e\ket{0}_n$, {\textbf{(upper)}}, the Lambda configuration showing the excited and ground states, and the polarization dependent resonant absorption process. Using the orbital angular momentum basis, $\ket{E_{0,\pm}}$, these excited and ground states are explicitly written as $\ket{A_1}=\frac{1}{\sqrt{2}}[\ket{E_-}\ket{+1}_e - \ket{E_+}\ket{-1}_e]$, $\ket{E_0}\otimes\ket{\pm 1}_e$ ~\cite{Maze_NV_structure}. {\textbf{(c)}}, The relaxation of the excited state $\ket{A_1}$ to the ground state $\ket{0}_e$ by a nonradiative decay, the single-shot readout of the electronic spin state by a continuous excitation to $\ket{E_x}$ and the detection of the emitted photons. The $\ket{0}_e \leftrightarrow \ket{E_x}$ cycling transition is used for both heralding and nuclear spin readout. {\textbf{(d)}}, Energy-level diagram for performing the nuclear spin state tomography. In the above figure the blue-solid arrows correspond to microwave transitions of the electronic spin state, green-solid arrows correspond to radio frequency transitions of the nuclear spin states and the red-solid arrows indicate optical transitions.
 }
\label{fig:1}
\end{figure*}
For this reason solid state devices with well controllable spins are recently proposed to be promising candidates for quantum repeater networks \cite{lukinqr,jiang}. 
The nitrogen-vacancy (NV) defect centre in diamond does show significant potential in this respect. It provides a hybrid spin system in which electron spins are used for fast~\cite{Awschalom_fast_control}, high-fidelity control~\cite{Doherty_review} and readout~\cite{Neumann_singleshot,Hanson_singleshot}, and nuclear spins which are well-isolated from their environment yielding ultra-long coherence time~\cite{Lukin_C13}. Electron and nuclear spins could form a small-scale quantum register allowing for e.g. necessary high-fidelity quantum error correction. Furthermore, the NV electron spin can be entangled with an emitted optical photon\cite{Lukin_photon,kosaka_entangled}. Quantum entanglement and quantum teleportation between two remote NV centres have already been experimentally demonstrated \cite{Hanson_entanglement,Hanson_teleportation}. A further and significant step towards using the NV centres in diamond to realize a functional quantum repeater network is to demonstrate its ability to store quantum information from a light field into the defects spins in such a way that it has the capability for a repetitive readin and readout of the memory, essential for scalable networks \cite{jiang}.

Quantum repeater networks can be realized either in the emission based approach where the photon entangled to a given solid-state qubit in a quantum node is absorbed by the other, thereby generating entanglement between adjacent nodes \cite{jiang}, or through an absorption based approach where absorption of a photon from an entangled pair between adjacent nodes could lead to entanglement between them.
While emission based approaches suffer from low absorption and collection efficiencies of photons from solid-state qubits and also a very low coherent emission into the Zero Phonon Line (ZPL), the later approach is only limited to low absorption efficiency of the solid-state system. Here, as the first step towards realising an absorption-based solid-state quantum repeater network we characterize a quantum node with solid-state spins in diamond that can be initialized, entangled and readout with very high efficiency. Relying only on photon absorption and spin measurements we show that the polarization states of the photon can be stored with high fidelity in a long-lived nuclear spin.


\begin{figure*}
\includegraphics[width=0.85\textwidth]{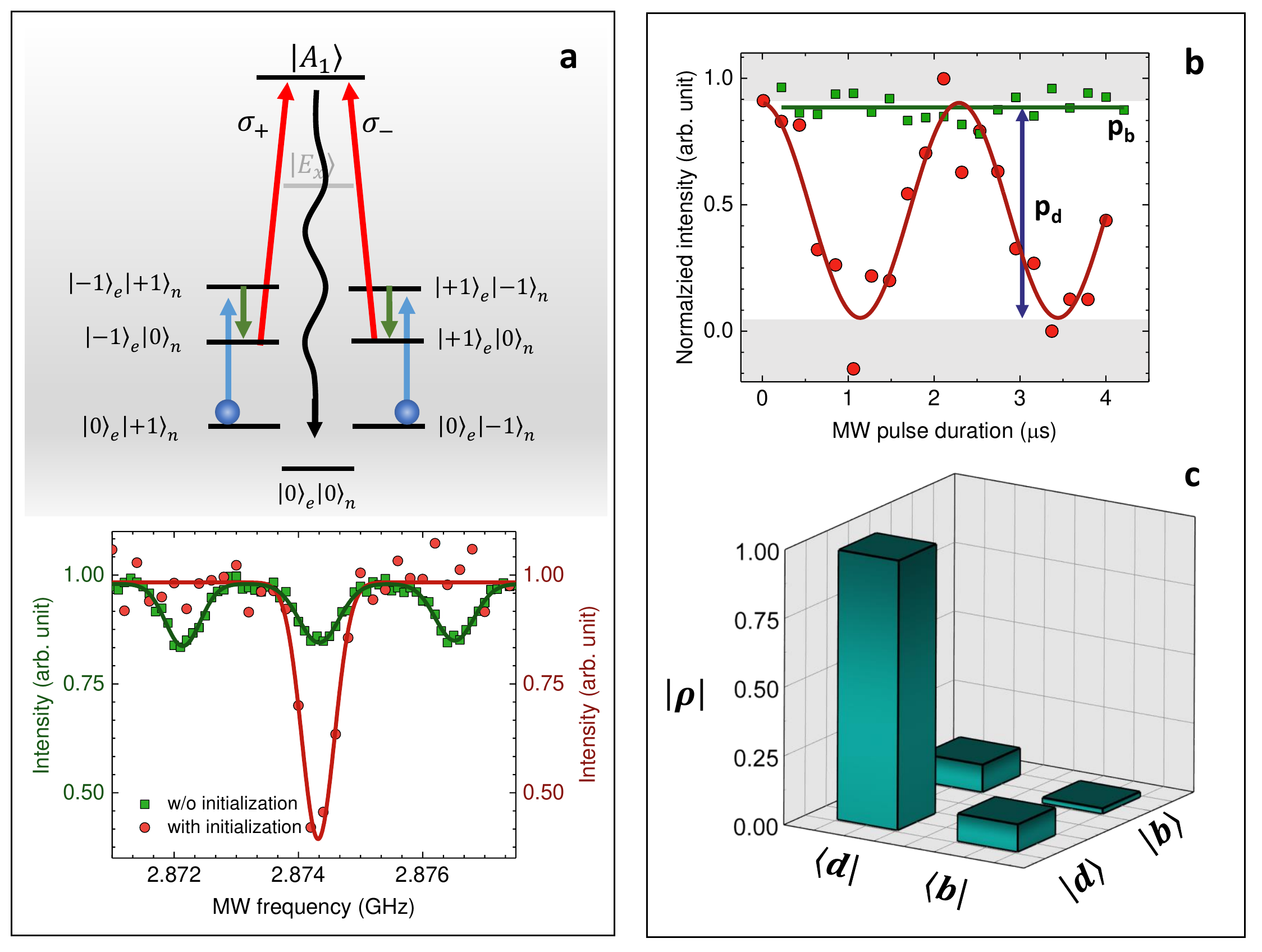}
	\caption{\textbf{Nuclear spin readout.}
	\textbf{(a)}, Deterministic initialization of \Niso nuclear spin.\textbf{(upper)}, Energy-level diagram describing the initialization of \Niso nuclear spin. The color code for various transitions involved are same to that given in Fig. 1. \textbf{(lower),} nuclear spin population measurements before (green) and after (red) nuclear initialization. The two different scales shown in the figure correspond to different methods involved in the measurement procedure. For measuring the nuclear spin population in its thermal state (green) we use the Optically detected Magnetic resonance (ODMR) of the electron spin and for measuring the nuclear spin population after initialization (red) we perform a single-shot readout of the electron spin state $\ket{0}_e$.
   \textbf{(b)}, Nuclear spin state measurement after storing photonic state $\frac{1}{\sqrt{2}}[\ket{\sigma_+}+\ket{\sigma_-}]$ . We measure the electron spin Rabi oscillation conditioned on the nuclear spin states $\ket{b}_n = \frac{1}{\sqrt{2}}[\ket{+1}_n + \ket{-1}_n]$(green) and $\ket{d}_n= \frac{1}{\sqrt{2}}[\ket{+1}_n - \ket{-1}_n]$(red), to obtain the nuclear spin state populations $p_b$ and $p_d$. The stored phase is then found from the simple equation, $\phi = \cos^{-1}[(p_b-p_d)/(p_b+p_d)]$, which in the present case is zero. Each data point is accumulated for 350 rounds.
    \textbf{(c)}, Reconstructed density matrix of nuclear spin from data in (b).
				}
\label{fig:2}
\end{figure*}

The basic element of our system is a single NV centre consisting of an electronic spin (S=1) and intrinsic \Niso nuclear
spin (I=1), coupled by hyperfine interaction. The selection rules for optical excitation allow both electronic spin states $\ket{\pm 1}_e$ to be excited to $\ket{A_1}$ through absorption of a photon with $\sigma_{+}$ and $\sigma_{-}$ polarization respectively, forming a Lambda system as shown in Fig. 1(a). Due to the short life time of the excited state the electron quickly relaxes to the ground states, $\ket{+1}_e,~\ket{-1}_e,~\ket{0}_e$  either by spontaneous emission within 12ns~\cite{Lukin_A1}, or through a combination of intersystem crossing and phonon or infrared photon emission within 250ns. As the photon is emitted accompanied by the excitation of fast decaying local phonons, the electron spin quickly dephases, loosing all its coherent information. Only in $4\%$ cases where the emission into ZPL happens, the electron remains entangled with the emitted photon.

The whole transfer and storage process and its verification consists of four steps:
Bell state preparation of the electron-nuclear spin system (Fig. 1(b) (lower)), optical writing (Fig. 1(b) (upper), 1(c)), heralding (Fig. 1(c)) and the nuclear spin readout (Fig. 1(d)). We start with the two electron spin ground states being degenerate and the NV spin system prepared in an initial Bell state $\ket{\Psi^{+}}=\frac{1}{\sqrt{2}}(\ket{+1}_e\ket{-1}_n + \ket{-1}_e\ket{+1}_n)$ . A photon in state $\ket{\psi}_p = \frac{1}{\sqrt{2}}(\ket{\sigma_+}+e^{i\phi}\ket{\sigma_-})$ and in resonance with the $A_1$ transition is sent into the NV centre. After absorption of a photon, the collective photon-NV spin system evolves into the  state $\ket{A_1}\otimes\ket{\psi}_n$ (see SOM), where $\ket{\psi}_n=\frac{1}{\sqrt{2}}(\ket{+1}_n - e^{i\phi}\ket{-1}_n)$. As our protocol depends only on successful absorption of a photon but not on the emission we are not limited by the $4\%$ ZPL. The successful storage of photon on the nuclear spin can be heralded either by detecting the emitted photon or by a single-shot readout of the electronic spin state $\ket{0}_e$ (see Fig. 1(c)). In the later way the readout photons being different from the read-in photons makes the heralding process almost $100\%$ efficient. Relying on the heralding of the spontaneously emitted photons from $\ket{A_1}$ is less efficient due to the low collection efficiencies of the detector. The perfect degeneracy between the ground states in a Lambda configuration allows absorption $50\%$ of the time (see SOM). Further with only a $40\%$ probability of finding the electron in spin state $\ket{0}_e$ after relaxation, the efficiency of the storage process ideally reaches a value of $\approx 20\%$ in the single photon picture (i.e., absorption and emission of only one photon). As there is no direct interaction between the memory (nuclear spin) and the incoming photon, unlike other systems\cite{Rempe_atom_memory,Rempe_PRL}, scattering photons would not lead to photon storage in our system.

We implement this protocol in a low strain ($\approx$1.2 GHz) NV centre to suppress strain-induced effects, e.g. lowering the symmetry of the NV and altering the configuration of the excited state. For experiments a NV centre along [111] orientation is chosen. In the experiment the net magnetic field is set to zero to ensure degeneracy of the ground states. By working at liquid helium temperature (T$<$8 K) we can resolve optical transitions and enable resonant excitation to perform efficient initialization and projective high-fidelity single-shot readout on the electron spin~\cite{Hanson_singleshot}.
The upper limit of the nuclear spin coherence time, given by the electron spin's $T_1$ time which reaches minutes at low temperatures.
\begin{figure*}
\includegraphics[width=0.85\textwidth]{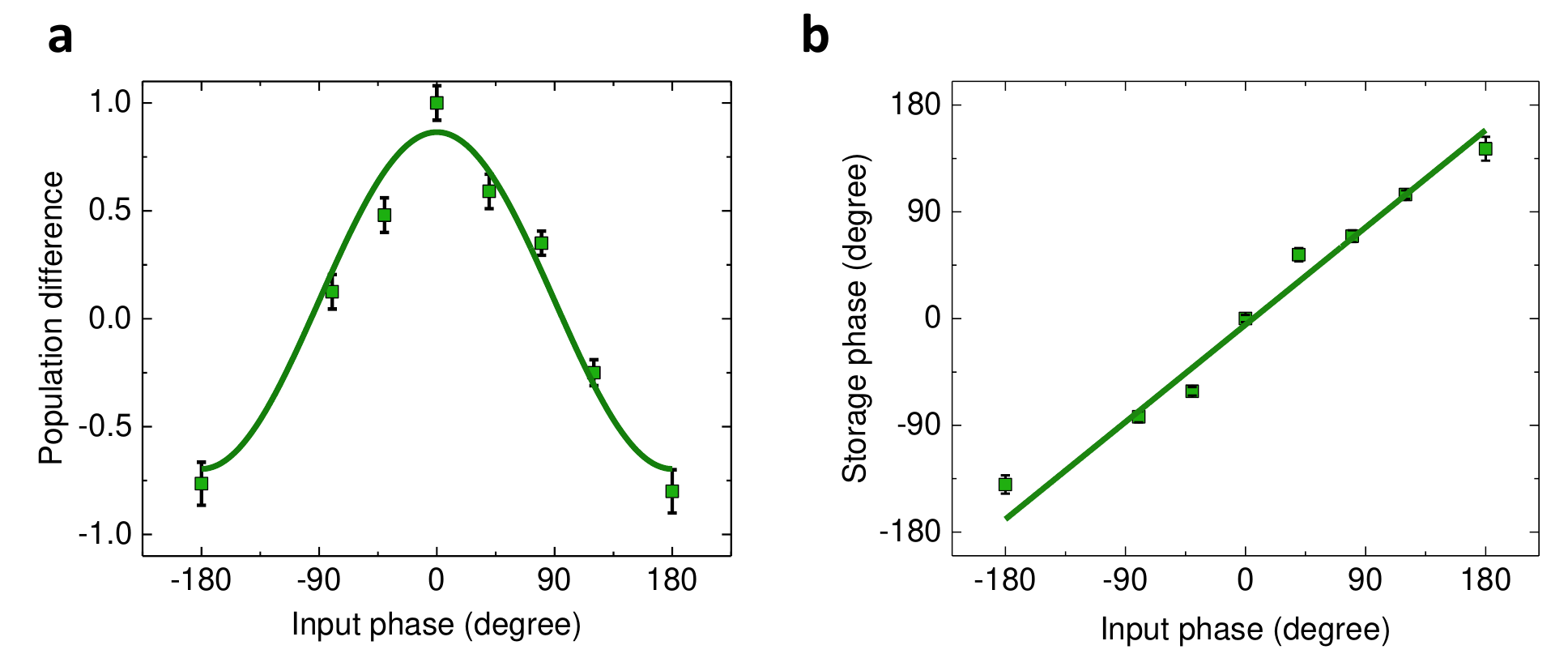}
	\caption{\textbf{Full phase space measurement.}
	\textbf{(a)}, Population difference between the dark and bright state of the nuclear spin state (defined in Fig.2), for different input phases of optical photons, which shows the $\cos\phi$ oscillation as expected.
   \textbf{(b)}, Comparison between the phase of the nuclear spin and the phase of optical photons.
		}
\label{fig:3}
\end{figure*}
Fig.~\ref{fig:1}(a) shows the circuit diagram and pulse sequence of the protocol.
Before preparing the Bell state, we first initialize the \Niso nuclear spin
into state $\ket{0}_n$ with a fidelity higher than $98\%$ (Fig.~\ref{fig:2}(a)) by applying polarization transfer from the electron spin. We achieve such high fidelity due to the initialization via the excited state $\ket{A_1}$ which is hardly effected by the hyperfine interaction.
To this end we start from state $\ket{0}_e$ of the electron spin and a thermal state of the \Niso nuclear spin characterized by identical populations
in states $\ket{0}_n$, $\ket{+1}_n$ and $\ket{-1}_n$.
Populations in states $\ket{0}_e\ket{+1}_n$ and $\ket{0}_e\ket{-1}_n$ are then transferred to state $\ket{+1}_e\ket{+0}_n$ and $\ket{-1}_e\ket{0}_n$ through
a nuclear-spin-controlled NOT gate followed by an electron-spin-controlled NOT gate on nuclear spin. After reinitializing the electron spin into state $\ket{0}_e$ , the final state is $\ket{0}_e\ket{0}_n$ (Fig.~\ref{fig:2}(a)).
{Unlike existing schemes, our approach here neither relies on level anti-crossings\cite{DNP_room} nor requires post-selection measurements~\cite{Hanson_singleshot}}.
The Bell state $\ket{\Psi^{+}}$ is prepared from $\ket{0}_e\ket{0}_n$ by applying a nuclear $\pi$ pulse followed by a nuclear-spin-controlled NOT gate (Fig. 1(b)).

In the second step after the Bell state is created, the electron in resonantly excited to  $\ket{A_1}$ by the photon to be stored. We synchronize the excitation to be within $20$ns after the Bell-state preparation to avoid any decoherence by the surrounding spin bath.  Here we use a laser pulse with a power of 200 nW (corresponding to an optical Rabi frequency of $27$MHz \cite{hansonrabi}) and pulse width $12$ns that is equal to the life time of the excited state, which contains a few thousand photons. Though we use a classical light field to demonstrate state conversion the defect can only absorb and emit single photons. The probability for absorbing more than one photon due to re-excitation is very small ($\sim0.025$, see SOM). The NV centre thus interacts with only a single photon during this $12$ns.

In the third step to confirm the absorption and subsequent relaxation of the electronic spin to its ground state we perform single-shot readout on the electron spin. First we map the phase information of the nuclear spin state on to its population by an RF-pulse (see Fig. 1(a)) i.e., $\frac{1}{\sqrt{2}}[\ket{+1}_n -{\rm e}^{i\phi} \ket{-1}_n] \rightarrow \cos\frac{\phi}{2}\ket{0}_n -i\sin\frac{\phi}{2} \ket{d}_n$. By doing so the phase can still be readout even if the nuclear spin dephases during the heralding of the electronic spin state. With an additional laser source we now induce cycling transitions, $\ket{0}_e \leftrightarrow \ket{E_{x,y}}$ (see Fig. 1(c)). By detecting the emitted photon flux we infer the electronic spin state which in turn confirms the storage process.

In the fourth and final step we readout the nuclear spin state when the electron spin is in state $\ket{0}_e$ corresponding to successful storage. By measuring the amplitude of electron spin Rabi oscillation conditioned on the nuclear spin state $\ket{0}_n$ and $\ket{\pm1}_n$, we obtain the populations of nuclear spin state, from which we infer the phase $\phi$. In Fig.~\ref{fig:2}(b) we show the Rabi oscillations of the electronic spin for $\phi=0$, and obtain the density matrix of the corresponding nuclear spin state which is shown in Fig.~\ref{fig:2}(c). Repeating the experiment with varying $\phi$ we readout the coherent phase between the photon polarization states via the nuclear spin state with an average fidelity of $98\%$(see Fig. 3).
 \begin{figure*}
\includegraphics[width=1.0\textwidth]{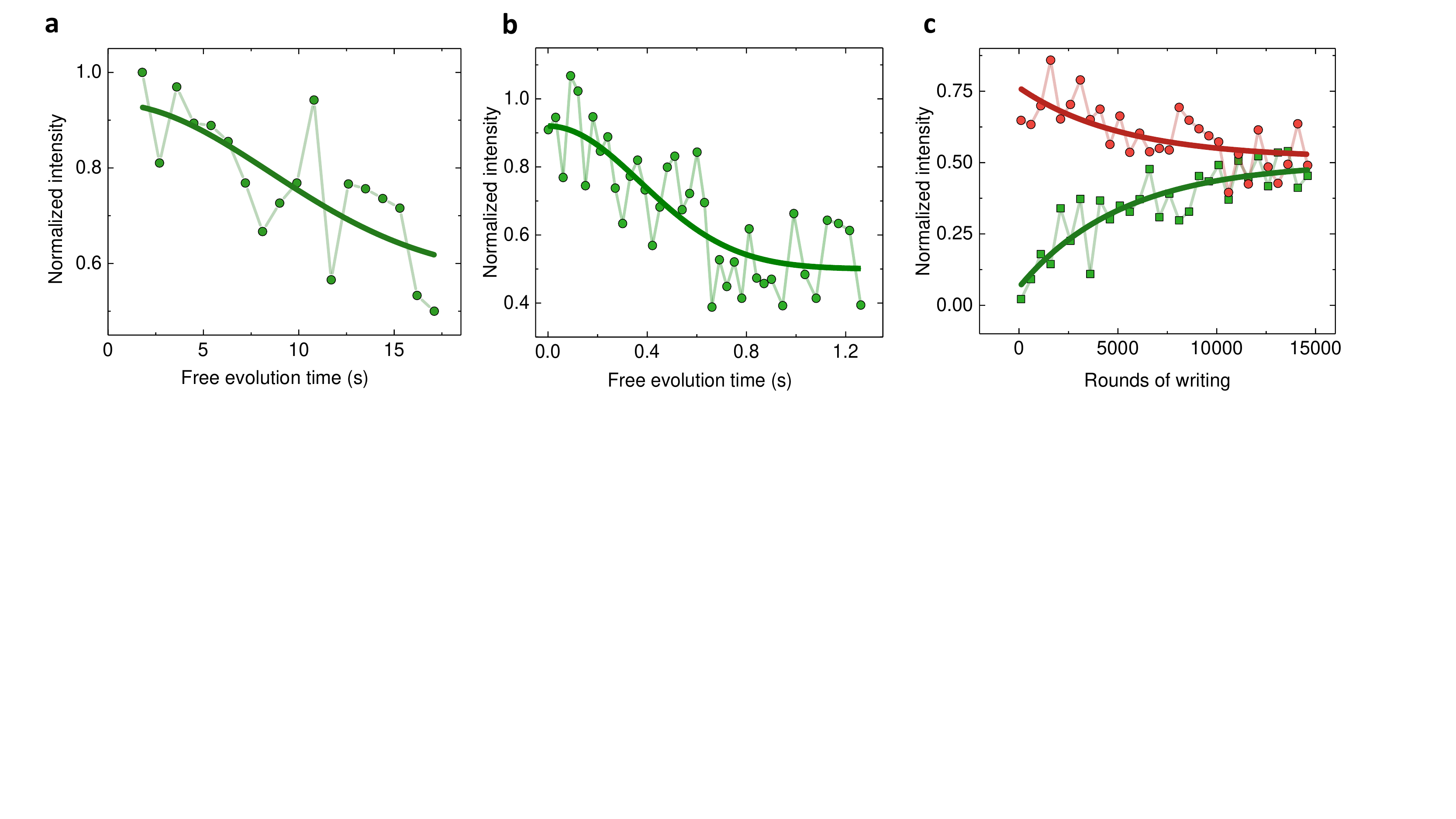}
	\caption{\textbf{Coherence of the nuclear spin memory.}
	\textbf{(a)}, Nuclear spin Hahn measurement for spin coherence between state $\ket{0}_e\ket{+1}_n$ and state $\ket{0}_e\ket{-1}_n$.
	\textbf{(b)}, Nuclear spin Hahn measurement for spin coherence between state $\ket{\pm1}_e\ket{\mp1}_n$ and state $\ket{\pm1}_e\ket{0}_n$.
	\textbf{(c)},  Nuclear spin Ramsey experiment with continuous $A_1$ laser illumination. Nuclear spin is
	prepared in state $\ket{0}_e(\ket{+1}_n+\ket{-1}_n)$ (green) and state $\ket{0}_e(\ket{+1}_n-\ket{-1}_n)$ (red).
		}
\label{fig:4}
\end{figure*}

We now verify the storage time of the memory, a key characteristic of a quantum node. Nuclear spins are considered to be a good quantum memory due to their natural isolation from the noisy environment.
Here we measure the coherence time of our $^{14}$N nuclear spin memory and study the decoherence mechanism. The free precession of the nuclear spin is measured via a Ramsey sequence (see SOM). The coherence dephases on a time scale of $T_{2n}^*= 0.31 \pm 0.05$s for the electron spin in $m_s=0$, and on a time scale of $T_{2n}^*= 12.0 \pm 1.9$ms for the electron spin in $m_s=\pm 1$. The dephasing arises from the noise generated by the surrounding $^{13}$C nuclear spin bath. The dependence on the electron spin state can be understood by noting the different spin bath dynamics in these two cases: for $m_s=0$ the bath's dynamics is dominated by nuclear magnetic dipole-dipole interaction, while the electron spin induced gradient magnetic field dominates in the $m_s=\pm 1$ case (see SOM).

To extend the coherence time, we apply a Hahn echo sequence. Fig.~\ref{fig:4}(a,b) shows the signal decay with a much longer coherence time. For the electron spin in the manifold $m_s=\pm 1$ the gradient magnetic field induced energy mismatch strongly suppresses flip-flop processes for the $^{13}$C nuclear spins close to the electron spin and should lead to a coherence time exceeding 5s (see SOM), which is in disagreement with the experimentally observed coherence time $T_{2n} = 0.53 \pm 0.09$s. This large discrepancy indicates that noise sources other than the nuclear spin bath is limiting spin coherence time in this case. In the $m_s=0$ manifold, however, such a limitation is not observed. A spin coherence time beyond 10s is measured, which can still be explained by treating the nuclear spin bath effects as an effective random classical field (see SOM). By comparing these two cases, the most probable factors causing the deviation in the $m_s=\pm1$ case would be the fluctuation of the hyperfine interaction between the NV
centre electron spin and the $^{14}$N nuclear spin, or the fluctuation of axial quadruple interaction of the $^{14}$N nuclear spin.

For a versatile quantum node that can perform entanglement distribution, entanglement distillation and quantum error corrections, multiple storage in different memory qubits (nuclear spins) are usually required. Hence the memory qubit should be robust against repetitive excitation of the communicator qubit. To verify this stability, we measure the nuclear spin's coherence time under continuous optical $A_1$ illumination. A coherence time $\sim 60\mu$s (Fig.~\ref{fig:4}(c)) which supports 5000 iterations of the optical writing is observed. The origin of the decoherence here can be understood by noting that the electron has a low but finite probability to be excited to state $\ket{E_{x,y}}$ due to non resonant optical illumination and subsequently relaxes back to the ground state $\ket{0}_e$ through spontaneous emission. In the excited state the spin-spin interaction and strain causes state mixture between state $\ket{E_{x,y}}$ and other excited states~\cite{Maze_NV_structure}, resulting in a new eigenstate $\ket{E^{'}_{x,y}}$. In this mixed state $\ket{E^{'}_{x,y}}$, the nuclear spin evolves under a Hamiltonian $H_e=\textbf{\textit{A}}\cdot\textbf{I}$ due to the hyperfine interaction, where
$\textbf{\textit{A}}=A_e(\left\langle E^{'}_{x,y}\right|\hat{S}_x\left|E^{'}_{x,y}\right\rangle, \left\langle E^{'}_{x,y}\right|\hat{S}_y\left|E^{'}_{x,y}\right\rangle, 0)$, $A_e\sim 40$MHz. A slow flipping induced by non resonant optical illumination is sufficient to dephase the nuclear spin (see SOM). To suppress this dephasing one can use a weakly coupled $^{13}$C spin instead of the $^{14}$N nuclear spin to reduce $A_e$.

In conclusion we show here that the absorption and heralded storage of a
photon (polarization states) in a nuclear spin can be achieved with high
fidelity using a combination of spin-spin, spin-photon Bell state
measurements and single-shot readout of electronic spin states. Bringing
in additional nuclear spins into action one can also implement
entanglement purification between the nuclear spins in two distant nodes
\cite{jiang} making the protocol fault-tolerant and also build other kinds
of quantum networks where a secure client driven computation can be
performed at quantum nodes that are blind to the input information. The photon storage at each node can also be further improved
by employing dynamical decoupling noise spectroscopy or a $^{12}$C
enriched diamond \cite{Enrich_diamond} for a longer electron spin
coherence times and using weakly coupled $^{13}$C nuclear spins for
increased stability of memory qubits under repetitive excitation.

The demonstrated results mark a first step towards realizing an
absorption based quantum repeater network using solid-state spins in
diamond as quantum nodes. As connecting these nodes require entangled
photon sources at $637$nm, the possibility of using NV centres in diamond
itself as a source to generate such photons could make these systems
stand-alone components in quantum repeater networks. Quite remarkably our scheme also lends itself for generating entangled photons. Entanglement
among the outgoing (emitted) photons is generated by repetitively entangling the
electron with the emitted photon and transferring it to the nuclear spin
\cite{durgaprb}. In this scheme nuclear spins also serve as quantum memory with coherence not destroyed by the interaction of the electron spin with photons. With the demonstrated stability of the nuclear spin under
repetitive excitation ($5000$ rounds) of the electron we estimate that a
$10$-photon GHZ state can be generated at a rate of 1Hz in the presence of
a cavity. Certainly, such an entangled photon source can also be used in linear optics quantum computing.

\bibliography{references}

\textbf{Acknowledgments} We thank Helmut Fedder, Ilja Gerhardt, Ingmar Jakobi, Kim Kafenda \& Kangwei Xia for technical supports, Marcus Doherty, Johannes Greiner, Renbao Liu, Fazhan Shi, Petr Siyushev \& Nan Zhao for fruitful discussions. M.M. and D.J.T. acknowledge finical support from The DARPA SPARQC program. H.K. ~acknowledges financial support by the the NICT Quantum Repeater Project. J.W. ~acknowledges financial support by the ERC project SQUTEC, the DFG SFB/TR21, the EU projects DIAMANT, SIQS and QESSENCE, JST-DFG (FOR1482), as well as the Volkswagenstiftung.
\\
\\
\textbf{Author Contributions} H.K. conceived the original idea, S.Y., Y.W. and P.N. designed the experiment,
S.Y. and T.H.T. performed the experiment, S.Y., Y.W. and D.D.B.
analyzed data and wrote the paper, J.W. supervised the project and all authors commented on the manuscript.
\\
\\
\textbf{Author Information} Reprints and permissions information is available at www.nature.com/reprints.
The authors declare no competing financial interests.
Readers are welcome to comment on the online version of the paper.
Correspondence and requests for materials should be addressed to to S.Y.(s.yang@physik.uni-stuttgart.de) 
\& J.W.(wrachtrup@physik.uni-stuttgart.de).
\end{document}